\documentclass{article}

\usepackage[T1]{fontenc}
\usepackage[utf8]{inputenc}
\usepackage{graphicx}
\usepackage{amsmath,amssymb}
\usepackage{booktabs}
\usepackage{float}     % for [H]
\usepackage[normalem]{ulem} % for \sout etc. without changing \emph
\usepackage{algorithm}
\usepackage{algpseudocode}
\usepackage{url}
\usepackage{hyperref}
\usepackage[dvipsnames]{xcolor}
\hypersetup{hidelinks}
\usepackage{stmaryrd}
\usepackage{paralist}
\usepackage{mathrsfs}
\usepackage{tikz}
\usetikzlibrary{shapes.geometric,arrows.meta}
\tikzset{
  >={Stealth[length=3.2mm,width=2.2mm]},
  every path/.append style={line width=0.6pt, shorten >=1pt}
}
\usepackage{pdflscape}
\usepackage{graphicx}

\newcommand{\zero}{\textsf{0}}
\newcommand{\one}{\textsf{1}}
\newcommand{\R}{\mathbb{R}}
\newcommand{\N}{\mathbb{N}}

\newcommand{\B}{\mathbb{B}}
\newcommand{\bool}{\{\zero,\one\}}
\newcommand{\integers}{\llbracket n \rrbracket}
\newcommand{\Inst}{\textit{inst}}
\newcommand{\Frus}{\textit{frus}}
\newcommand{\pcycle}{C^+}
\newcommand{\ncycle}{C^-}
\newcommand{\pcomp}{\textit{c}^+}
\newcommand{\ncomp}{\textit{c}^-}
\newcommand{\traj}{\mathcal{T}\hspace*{-3pt}\textit{raj}}
\newcommand{\sens}{\mathcal{S}\hspace*{-1.7pt}\textit{ens}}
\newcommand{\miR}{mi\textsc{r}200}
\newcommand{\zeb}{\textsc{zeb}}
\newcommand{\spone}{\textsc{sp}1}
\newcommand{\smad}{\textsc{smad}3}
\newcommand{\tgfb}{\ensuremath{\textsc{tgf}\beta}}
\newcommand{\notch}{\textsc{notch}1}
\newcommand{\col}{\textsc{col}1\textsc{a}1}
\newcommand{\ccnd}{\textsc{ccnd}1}
\newcommand{\pax}{\textsc{pax}8}

\begin{document}

\title{Parsimonious computational inference protocol for Boolean networks: Application to osteogenesis}
	
\author{\normalfont Jacques Demongeot\,$^{1,}$\thanks{
		\href{mailto:jacques.demongeot@univ-grenoble-alpes.fr}{jacques.demongeot@univ-grenoble-alpes.fr}
	}~, Alonso Espinoza Rojas\,$^{2,3,}$\thanks{
	  \href{mailto:alonso.espinoza-rojas@lis-lab.fr}{alonso.espinoza-rojas@lis-lab.fr}
	}~, Eric Goles\,$^{2,4,}$\thanks{
	  \href{mailto:eric.chacc@uai.cl}{eric.chacc@uai.cl}
	}~,\\ 
	Marco Montalva-Medel\,$^{2,}$\thanks{
	  \href{mailto:marco.montalva@uai.cl}{marco.montalva@uai.cl}
	}~, Sylvain Sen{\'e}\,$^{3,6,}$\thanks{
		\href{mailto:sylvain.sene@lis-lab.fr}{sylvain.sene@lis-lab.fr}
	}~, Laurent Tichit\,$^{5,6,}$\thanks{
	  \href{mailto:laurent.tichit@univ-amu.fr}{laurent.tichit@univ-amu.fr}
	}\\[2mm]
	{\small $^1$~Universit{\'e} Grenoble Alpes, La Tronche, France}\\
	{\small $^2$~Universidad Adolfo Ib{\'a}{\~n}ez, Pe{\~n}alol{\'e}n, Santiago, Chile}\\
	{\small $^3$~Aix Marseille Univ, CNRS, LIS, Marseille, France}\\
	{\small $^4$~Millennium Nucleus for Social Data Science, Santiago, Chile}\\
	{\small $^5$~Aix Marseille Univ, CNRS, I2M, Marseille, France}\\
	{\small $^6$~Universit{\'e} publique, Marseille, France}
}

\date{}

\maketitle

\begin{abstract}
Boolean networks are powerful mathematical tools for modeling the qualitative dynamics 
of genetic regulation. 
Yet inferred models often generate spurious attractors that lack biological viability. 
In this paper, we propose a parsimonious computational framework to systematically 
refine Boolean network models by eliminating these non-biological asymptotic behaviors 
while strictly preserving known, biologically relevant attractors. 
Through an exhaustive exploration of local function substitutions, we generate a 
comprehensive set of candidate models. 
To identify the most biologically consistent networks, we implement an incremental 
pruning protocol that filters candidates based on structural interaction digraph 
similarity, attraction basin topological organization, trajectorial isomorphism, and 
the minimization of dynamical instability and frustration. 
We apply this methodology to a 9-node genetic control model of the osteogenesis 
regulation network. 
Our protocol effectively evaluates a syntactic search space of 51,138 potential 
networks, ultimately narrowing them down to a robust family of 6 parsimonious models 
that are fully compatible with current biological knowledge.\\[2mm]
\emph{Keywords: Biological system modelling \and Boolean networks \and 
    Discrete dynamical systems.} 
\end{abstract}
\setcounter{footnote}{0}
\section{Introduction}

The study of finite discrete dynamical systems, and particularly of Boolean networks, 
has been a fundamental pillar for understanding information propagation and the 
dynamics of complex systems~\cite{goles1990neural}. 
Mathematically, a finite discrete dynamical system is governed by a finite 
configuration space that evolves over discrete time steps through a global transition 
function~\cite{robert1986discrete}. 
Boolean networks represent a specific and highly effective subset of these models. 
In a Boolean network, the model is defined by a collection of local Boolean functions, 
from which a directed interaction graph of $n$ nodes is derived. 
Each node holds a local binary state from $\{0, 1\}$. 
A configuration of the network belongs to the configuration space $\{0, 1\}^n$, and 
the temporal evolution of each local state is dictated by a specific Boolean function 
depending on the states of its interacting neighbors.

Originally introduced in the context of theoretical neural networks by McCulloch and 
Pitts~\cite{mcculloch1943logical}, Boolean networks quickly demonstrated their immense 
utility in the field of theoretical biology. 
Pioneers such as Kauffman and Thomas successfully adapted these networks to model 
genetic regulation networks, laying the foundations for the qualitative representation 
of biological interactions~\cite{kauffman1969metabolic,thomas1973boolean}.

In this work, we situate ourselves at the interface between discrete mathematics, 
computer science, and theoretical biology, developing a general parsimonious 
algorithmic method to infer more likely regulation models from a given one, and 
applying it to a specific genetic control model: the osteogenesis regulation 
network~\cite{demongeot2025boolean}. 
More precisely, basing ourselves on this network as a case study, our main objective 
is to transform this initial model by systematically eliminating \emph{spurious 
attractors}. 
Spurious attractors are those limit cycles or fixed points generated by the 
mathematical model that lack an apparent biological meaning or viability in the 
literature. 
By exploring local rule substitutions, we generate a comprehensive set of candidate 
models designed to strictly preserve a prescribed set of biologically relevant fixed 
points while removing undesired asymptotic behaviors. 
To select the optimal models from this extensive set, we implement an incremental 
pruning protocol. 
This approach ensures biological conservation by systematically filtering candidates 
based on their structural interaction digraph similarities, the topological 
organization and relative sizes of their attraction basins, trajectorial isomorphism, 
and the minimization of dynamical frustrations and instabilities.

In Section~\ref{sec:theo}, we present the main definitions and notations which are 
used throughout the paper. 
Section~\ref{sec:proto} defines the parsimonious protocol allowing to obtain as a 
result a family of models complying with the initial hypotheses. 
In Section~\ref{sec:appli}, the protocol is applied to the osteogenesis control 
network mentioned above, which leads to obtain a family of six more likely Boolean 
models.
In conclusion, we give some perspectives for this work.

\section{Theoretical concepts}
\label{sec:theo}

Given $n \in \N$, we note $\integers = \{0, \ldots, n-1\}$.
Let $\B = \bool$. 
For $x \in \B^n$,  and $i \in \integers$, we denote by $x_i$ the $i$th component of 
$x$, and by $x + e_i$ the vector of $\B^n$ obtained by flipping it (addition modulo 
$2$). 
Let $x = (x_0, \ldots, x_{n-1}) \in \B^n$, we also denote it as the word $x_0 \ldots 
x_{n-1}$ for the sake of simplicity.
We refer to the concept of a directed cycle in graph theory simply as a \emph{cycle}.

\paragraph{Boolean networks.}
\label{par:bn}

A \emph{Boolean network} (BN) $f$ of size $n$ is a function mapping $\B^n$ to itself. 
We see $f$ as a collection of $n$ \emph{local functions} $(f_0, \dots, f_{n-1})$, such 
that for all $i \in \integers$, $f_i: \B^n \to \B$.
In what follows, $x \in \B^n$ (resp. $x_i \in \B$) represents a \emph{configuration} 
(resp. the \emph{state} of component $i$).\smallskip 

\begin{figure}[t!]
	\centerline{
		\begin{minipage}{.32\textwidth}
			\centerline{$f(x) = \left(\begin{array}{l}
				f_0(x) = x_1\\
				f_1(x) = x_0 \land x_2\\
				f_2(x) = \neg x_0 \lor x_1\\
			\end{array}\right)$}
		\end{minipage}
		\qquad\qquad\qquad
		% Columna derecha: Gráfico de influencia
		\begin{minipage}{.18\textwidth}
			\centerline{\scalebox{.85}{\begin{tikzpicture}[>=latex,auto]
				\tikzstyle{node} = [circle, thick, draw]
				\node[node](n0) at (0,0) {$0$};
				\node[node](n1) at (-1,-1.5) {$1$};
				\node[node](n2) at (1,-1.5) {$2$};
				
				% Interacciones entre a y b
				\draw[thick, ->] (n0) edge [bend right=20] node [swap] {$+$} (n1);
				\draw[thick, ->] (n1) edge node [swap] {$+$} (n0);
				
				% Interacción de a hacia c (negativa)
				\draw[thick, ->] (n0) edge node {$-$} (n2);

				% Interacciones entre b y c
				\draw[thick, ->] (n1) edge [bend right=20] node [swap] {$+$} (n2);
				\draw[thick, ->] (n2) edge node [swap] {$+$} (n1);
			\end{tikzpicture}}}
    	\end{minipage}
    }
	\caption{Example of representation of a BN $f$ of size $n=3$. 
    (Left) The local functions defining the network. 
    (Right) Its corresponding signed interaction digraph $G_f$.	
	\label{fig:1example}}
\end{figure}
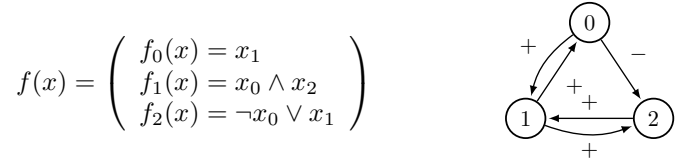

The \emph{interaction digraph} of a BN $f$ represents the effective dependencies 
between its components, and is defined as $G_f = (\integers, E_f)$, where $E_f$ is the 
set or arcs such that
\begin{equation*}
    (i,j) \in E_f \iff
    \exists x \in \B^n,\, f_j(x) \neq f_j(x + e_i)\text{.}
\end{equation*}
In this paper, we only focus on simple interaction 
digraphs and not on multidigraphs, 
which leads us to focus  only on locally monotonic BNs.
Arcs of interaction digraphs may be assigned signs $\sigma: E_f \to \{-,+\}$ as 
follows:
\begin{itemize}
  \item $\sigma(i,j)=+$ when $\exists x \in \bool^n : x_i=1 \wedge 
  f_j(x) > f_j(x+e_i)$,
  \item $\sigma(i,j)=-$ when $\exists x \in \bool^n : x_i=1 \wedge 
  f_j(x) < f_j(x+e_i)$.
\end{itemize}
A cycle in $G_f$ is \emph{positive} (resp. \emph{negative}) if the number of its 
negative arcs is even (resp. odd).
%\textcolor{OliveGreen}{For the thesis, we should also define these cycles as 
%"structural circuits", and define "functional circuits" cf R. Thomas (circuits where 
%all the interactions are functional in a common set of states). These are the ones 
%that are useful (and linked to R. Thomas conjectures). GINsim has a function 
%enumerating functional circuits, bioLQM can be used to script the same functionality}
An example of a BN of size $n = 3$ and its corresponding interaction digraph is 
provided in Figure~\ref{fig:1example}. 
The signs of the directed arcs directly reflect the monotonic dependencies induced by 
the local functions, such as the negative influence of component $0$ on component $2$ 
resulting from the negation $\neg x_0$ in $f_2$.

\paragraph{Update modes and discrete dynamical systems.}
\label{par:um_dds}

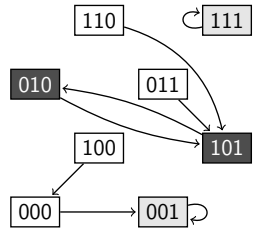
\begin{figure}[t!]
	\centerline{
		\begin{minipage}{.4\textwidth}
		    \centerline{\begin{tabular}{@{}c|c|c|c|c@{}}
                \hline
		          $x = x_a x_b x_c$ &
		          $f_a(x)$ & 
	            $f_b(x)$ & 
		          $f_c(x)$ & 
		          $f(x)$\\
	            \hline
		          $\zero\zero\zero$ & $\zero$ & $\zero$ & $\one$ & $\zero\zero\one$\\
		        $\zero\zero\one$ & $\zero$ & $\zero$ & $\one$ & $\zero\zero\one$\\
		          $\zero\one\zero$ & $\one$ & $\zero$ & $\one$ & $\one\zero\one$\\
		          $\zero\one\one$ & $\one$ & $\zero$ & $\one$ & $\one\zero\one$\\
		        $\one\zero\zero$ & $\zero$ & $\zero$ & $\zero$ & $\zero\zero\zero$\\
		          $\one\zero\one$ & $\zero$ & $\one$ & $\zero$ & $\zero\one\zero$\\
		          $\one\one\zero$ & $\one$ & $\zero$ & $\one$ & $\one\zero\one$\\
		          $\one\one\one$ & $\one$ & $\one$ & $\one$ & $\one\one\one$\\
		          \hline
		      \end{tabular}}
		\end{minipage}
        \qquad\qquad\qquad
        \begin{minipage}{.26\textwidth}
            \centerline{\scalebox{.85}{\begin{tikzpicture}[>=to,auto]
		        \tikzstyle{type} = []
		        \tikzstyle{conf} = [rectangle, draw]
		        \tikzstyle{pf} = [rectangle, draw, fill=black!10]
	            \tikzstyle{lc} = [rectangle, draw, fill=black!70]
                \node[conf](n000) at (0,0) {$\zero\zero\zero$};
		        \node[pf](n001) at (2,0) {$\zero\zero\one$};
		        \node[lc](n010) at (0,2) {\textcolor{white}{$\zero\one\zero$}};
	            \node[conf](n011) at (2,2) {$\zero\one\one$};
		        \node[conf](n100) at (1,1) {$\one\zero\zero$};
		        \node[lc](n101) at (3,1) {\textcolor{white}{$\one\zero\one$}};
		        \node[conf](n110) at (1,3) {$\one\one\zero$};
	            \node[pf](n111) at (3,3) {$\one\one\one$};
                \draw[->] (n000) edge (n001);
		        \draw[->] (n001) edge [loop right, distance=4mm] (n001);
		        \draw[->] (n010) edge [bend right=10] (n101);
	            \draw[->] (n011) edge (n101);
		        \draw[->] (n100) edge (n000);
		        \draw[->] (n101) edge [bend right=10] (n010);
		        \draw[->] (n110) edge [bend left=30] (n101);
	            \draw[->] (n111) edge [loop left, distance=4mm](n111);
		    \end{tikzpicture}}}
        \end{minipage}
	}
	\caption{(Left panel) Configurations, local updates, and their images according to 
        the parallel update mode for the BN $f$ introduced in 
        Figure~\ref{fig:1example}. (Right panel) The associated transition graph 
        $\mathscr{G}_f$.}
    \label{fig:updates_example}
\end{figure}

Given a BN $f$, a configuration $x$ and a subset $I \subseteq \integers$, we denote 
by $f_{I}(x)$ the configuration obtained by the \emph{update} of components of $I$ 
from $x$. 
Formally, we have:
\begin{equation*}
    \forall i \in \integers,\, f_I(x)_i = 
    \left\{\begin{array}{ll}
    f_i(x) & \text{ if } i \in I\\
    x_i & \text{ otherwise.}
  \end{array}\right.
\end{equation*}
Note that $f_{\integers}=f$.
A {\em block-sequential update mode} (BSUM) is an ordered partition of $\integers$, 
denoted by $\mu = (W_1, \ldots, W_p)$ (with $p$ the \emph{period} of $\mu$), and a BN 
$f$ updated according to $\mu$ gives the deterministic \emph{discrete dynamical 
system} (DDS) on $\B^n$ defined as:
\begin{equation*}
    f_\mu = f_{W_p} \circ \dots \circ f_{W_2} \circ f_{W_1}\text{.}
\end{equation*}
The update mode $(\integers)$ is called {\em parallel}.
Given $\mu = (W_1, \ldots, W_p)$ an ordered partition of $\integers$ and $x \in 
\B^n$, computing $f_\mu(x)$ corresponds to making $x$ evolve to its image over 
\emph{one time step}.
Moreover, given $k \in \{1, \ldots, p\}$, computing $f_{W_k} \circ \ldots \circ 
f_{W_1}(x)$ corresponds to computing the $k$th \emph{substep} of the evolution of $x$ 
within the period of $\mu$. 
Let $f_\mu$ be a DDS. 
It is represented by its functional digraph, classically called the \emph{transition 
graph}, defined as $\mathscr{G}_{f_\mu} = (\B^n, f_\mu: \B^n \to \B^n)$. 
Figure~\ref{fig:updates_example} presents the parallel DDS associated with the BN $f$ 
of Figure~\ref{fig:1example}. 
%The table details all possible configurations in $\mathbb{B}^3$ alongside the 
%resulting synchronous transition $f(x)$, which corresponds to a single time step of 
%the discrete dynamical system.

\paragraph{Trajectories, attractors and attraction basins.}
\label{par:attractors}

Let $f$ be a BN and $\mu$ be a BSUM. 
Given a configuration $x \in \B^n$, the \emph{trajectory} of $x$ is the infinite
path from $x$ in $\mathscr{G}_{f_\mu}$, defined as:
\begin{equation*}
    x = f_\mu^0(x) \to x^1 = f_\mu(x) \to x^2 = f_\mu(f_\mu(x)) = f_\mu^2(x) 
    \to \ldots 
\end{equation*}

Since the configuration space $\B^n$ is finite, every trajectory eventually enters a 
cycle of periodic configurations, called \emph{attractor}.
When the attractor consists of a single configuration (resp. several configurations), 
we speak of a \emph{fixed point} (resp. a \emph{limit cycle}) corresponding to 
a stationary (resp. oscillating) behavior.

Let $\mathcal{A}$ be an attractor of $f_\mu$. Its \textit{attraction basin} 
$B(\mathcal{A})$ is the set of configurations whose trajectories under $f_\mu$ 
eventually reach $\mathcal{A}$, such that:
\begin{equation*}
  B(\mathcal{A}) = \left\{ x \in \B^n \mid 
    \exists t_0 \in \N, \forall t \geq t_0,\, f_\mu^t(x) \in 
    \mathcal{A}\right\}.
\end{equation*}

Figure~\ref{fig:updates_example} highlights that the example network given in 
Figure~\ref{fig:1example} admits in parallel three attractors: two fixed points, 
$\zero\zero\one$ and $\one\one\one$ and one limit cycle of length $2$, 
$(\zero\one\zero, \one\zero\one)$, whose attraction basins are respectively 
$\{\zero\zero\zero, \zero\zero\one, \one\zero\zero\}$, $\{\one\one\one\}$, and 
$\{\zero\one\zero, \zero\one\one, \one\zero\one, \one\one\zero\}$ which can also be 
viewed as the respective induced subgraphs in the associated transition graph. 

\paragraph{Instability and frustration.}
\label{par:instability_frustration}

Let $f$ be a BN of size $n$, $G_f = (\integers, E_f)$ its associated interaction 
digraph, $\mu$ a BSUM on $\integers$, and $x \in \B^n$ a configuration. 
One may associate with $x$ its level of \emph{instability}~\cite{noual2018synchro}, 
defined as the function $\Inst$ that maps $\B^n$ to $\R$ such that:
\begin{equation*}
    \Inst(x) = d_H(x, f_\mu(x)) = \sum_{i = 0}^{n-1} (x_i - f_\mu(x)_i)^2\text{,}
\end{equation*} 
where $d_H(x, y)$ equals the Hamming distance between configurations $x$ and $y$.
It measures the amount of dynamical changes associated with the transition issued from 
$x$, and can be viewed as a discrete analogue of the kinetic energy based on 
configuration variations along its 
trajectory~\cite{demongeot2018entropy,rachdi2020entropy}.

One may also associate with $x$ its level of 
\emph{frustration}~\cite{toulouse1977frus,noual2018synchro}, 
defined as the function $\Frus$ that maps $B^n$ to $\N$ such that:
\begin{equation*}
    \Frus(x) = \sum_{i, j \in \integers} \Frus_{(i, j)}(x)\text{,}
\end{equation*}
where
\begin{equation*}
    \Frus_{(i, j)}(x) = 
        \begin{cases}
            1 & \text{if } (i,j) \in E_f \text{, } \sigma(i,j) = + \text{ and } 
                x_i \neq x_j\text{,}\\
            1 & \text{if } (i,j) \in E_f \text{, } \sigma(i,j) = - \text{ and } 
                x_i = x_j\text{,}\\
            0 & \text{if } (i,j) \notin E_f\text{.}
        \end{cases}
\end{equation*}
It quantifies to what extent a configuration is locally incompatible with the signs of 
the underlying interactions. 
It is a structural-dynamical observable that captures local inconsistencies between 
node states and their interactions.

\section{Parsimonious protocol}
\label{sec:proto}

Let $M_0$ denote an initial BN $f$ of size $n$, and let
$\mathcal{A}^* = \{\mathcal{A}_1, \dots, \mathcal{A}_k\}$ be a prescribed set of 
attractors selected \emph{a priori} as biologically relevant.
The aim of this algorithm is to generate from $M_0$ a family of alternative models 
obtained by local function substitutions, under the constraint that the resulting 
dynamics preserves exactly the prescribed set of biologically relevant attractors 
$\mathcal{A}^*$, while eliminating all remaining attractors, in particular spurious 
limit cycles and other non-selected asymptotic behavior. 
This places the procedure in a purely selective framework: 
the algorithm does not attempt to discover new relevant attractors, but rather to 
remove undesired ones while maintaining those of interest.

\subsection{Model generation}

Considering a BN $f$ of size $n$, we associate with each component $i\in 
\integers$ two possible literals, the positive one $x_i$ and the negative one 
$\neg x_i$.
The search space for candidate functions is defined in terms of literals rather than 
arbitrary Boolean functions. 
We consider candidate replacement functions involving at most three literals, where 
each variable appears at most once (due to the local 
monotonicity constraint), and thus contributes with to at most one literal. 
The literals are combined using the logical operators $\land$, $\lor$, and parentheses 
to form valid propositional logic formulae. 

\begin{table}[t!]
    \centerline{
        \begin{tabular}{c}
            \hline
            Syntactical structures\\
            \hline\hline
            $A \land B \land C$ \\
            $A \lor B \lor C$ \\
            $A \land B \lor C$ \\
            $A \land C \lor B$ \\
            $B \land C \lor A$ \\
            $(A \lor B) \land C$ \\
            $(A \lor C) \land B$ \\
            $(B \lor C) \land A$ \\
            \hline
        \end{tabular}
    }\bigskip
    
    \caption{Syntactical structures of a propositional formula of arity $3$ where $A$, 
        $B$ and $C$ are distinct positive or negative literals.}
    \label{tbl4}
\end{table}

For a fixed $k \in  \{1, 2, 3\}$ variables, the number of admissible 
functions is given by the following rules:
\begin{itemize}
\item When $k = 1$, two possible signs (\emph{i.e.} literals) give $2$ 
    distinct local functions.
\item When $k = 2$, $2^2$ sign assignments and $2$ operator choices give 
    $2^2 \cdot 2 = 8$ distinct local functions.
\item When $k = 3$, there are $2^3$ sign assignments and $8$ (parentheses/
    operators)-based admissible formulae (corresponding to distinct Boolean 
    functions), giving $2^3 \cdot 8 = 64$ distinct local functions 
    (cf. Table~\ref{tbl4}).
\end{itemize}
Therefore, for each component $i$, the number of replacement functions is:
\begin{equation*}
    |\mathcal{F}_i| = 2\binom{n}{1} + 8\binom{n}{2} + 64\binom{n}{3}.
\end{equation*}

We define by $M_{i,\varphi} = f^{(i,\varphi)} = 
(f_1,\dots,f_{i-1},\, \varphi,\, f_{i+1},\dots,f_n)$ an alternative model obtained by 
replacing the local function $f_i$ of node $i$ with a candidate function 
$\varphi \in \mathcal{F}_i$.
The full set of candidate models is $\mathcal{M} = 
\{ M_{i,\varphi} \mid i \in \integers,\ \varphi \in \mathcal{F}_i \}$
such that $|\mathcal{M}| = 
n\left(2\binom{n}{1} + 8\binom{n}{2} + 64\binom{n}{3}\right) = \Theta(n^4)$.
For example, $|\mathcal{M}|$ equals $282$ for $n = 3$, $1248$ for $n = 4$, ..., 
$51138$ for $n = 9$. 

\subsection{Model selection}

Once $\mathcal{M}$ has been built, the purpose of this stage is to perform an 
incremental pruning of $\mathcal{M}$ by retaining the models that remain the closest 
to the reference model $M_0$ with respect to selected structural and dynamical 
properties. 
Once a measure is computed, we remove from $\mathcal{M}$ a proportion $P \in [0\,;1]$ 
of the models among those which admit the largest values for the measure. 

%The first model selection step consists in removing from $\mathcal{M}$ all the models 
%whose attractor sets are not $\mathcal{A}^*$.
%Then, we introduce a collection of complementary similarity measures. 
%We first consider a structural criterion based on the comparison of interaction 
%digraphs, and then a family of dynamical criteria based on the organization of 
%attraction basins and on the associated convergence structures. 
%Taken together, these measures provide a systematic framework for selecting, within 
%$\mathcal{M}$, the models that best preserve the relevant features of $M_0$ while 
%satisfying the prescribed attractor constraints.

\subsubsection{Step 1: Attractor preservation}

Let $\mathcal{A}^* = \{\mathcal{A}_1, \ldots, \mathcal{A}_k\}$ be the prescribed set 
of attractors selected \emph{a priori} as biologically relevant.
Each candidate model $M_{i,\varphi}$ is retained if and only if its attractor set 
coincides exactly with $\mathcal{A}^*$. 
In practice, a candidate is first pre-filtered by checking that $\varphi$ reproduces 
the value of node $i$ at every recurring configuration of each prescribed attractor 
$\mathcal{A}_j \in \mathcal{A}^*$, with $j \in \{1, \ldots, k\}$; 
if so, the parallel dynamics of the candidate is simulated and the latter is conserved 
only if it does admit attractors only in $\mathcal{A}^*$.
The exhaustive exploration of all admissible substitutions then leads to the 
construction of the family $\mathcal{M} = \{ M_{i,\varphi} \in \mathcal{M}^\star \mid 
\mathrm{Attr}(M_{i,\varphi}) = \mathcal{A}^* \}$, where $\mathrm{Attr}(M_{i,\varphi})$ 
denotes the set of attractors of $M_{i,\varphi}$.

\subsubsection{Step 2: Interaction digraph similarity}

We introduce a structural similarity criterion based on the comparison of the 
interaction digraph of each model $M_m \in \mathcal{M}$ with that of the 
reference model $M_0$.
From the interaction digraph structural standpoint, it is well known that interaction 
cycles play an essential role on the dynamical behavior of 
BNs~\cite{robert1980thm,thomas1981conj,demongeot2012cycles}. 
So, being confident with the functional likeliness of $M_0$, the general idea is to 
conserve models of $\mathcal{M}$ whose interaction digraphs have similar structures 
with $M_0$ in terms of these cycles, and more precisely in terms of their number 
and the number of their components. 
Thus, let $\pcycle$ (resp. $\ncycle$) denote the number of positive (resp. negative) 
cycles, and $\pcomp$ (resp. $\ncomp$) denote the number of components belonging to 
positive (resp. negative) cycles of a given interaction digraph.

For $M_0$ and $M_m \in \mathcal{M}$, with respective interaction digraphs $G_0$ and 
$G_m$, the interaction digraph similarity index is defined as:
\begin{equation*}
    \Delta_G = \Delta(G_0, G_m) = 
        \left| \pcycle_{G_0} - \pcycle_{G_m} \right| +
        \left| \ncycle_{G_0} - \ncycle_{G_m} \right| + 
        \left| \pcomp_{G_0} - \pcomp_{G_m} \right| +
        \left| \ncomp_{G_0} - \ncomp_{G_m} \right|\text{.}
\end{equation*}

\subsubsection{Step 3: Dynamical similarities}

We introduce dynamical similarity criteria to compare the organization of attraction 
basins and convergence patterns between $M_0$ and the models of $\mathcal{M}$. 
Consider that $M_0$ has $\ell \in \N$ attractors and that $\ell - k$, with $k \leq 
\ell$, of them have been considered irrelevant from the biological standpoint. 
To do so, we denote the list of attraction basins of the parallel DDS of $M_0$ by 
$\mathcal{B}_0 = \left( B_1^{(0)}, \ldots, B_\ell^{(0)} \right)$, organized such that 
the basins associated with irrelevant attractors are all $B_i$s with 
$i \in \{k+1, \ldots, \ell\}$; 
and the list of attraction basins of $M_m \in \mathcal{M}$ by $\mathcal{B}_m = 
\left( B_1^{(m)}, \ldots, B_k^{(m)} \right)$, such that, for all 
$i \in \{1, \ldots, k\}$, the corresponding attractor of $B_i^{(0)}$ equals that of 
$B_i^{(m)}$.

\paragraph{Step 3.1: Relative size and attraction basin organization similarity.}

Let $B_i^{(0)}$ be an attraction basin of $M_0$ and $B_i^{(m)}$ be its associated 
attraction basin with $M_m$ (cf. surjection from $\mathcal{B}_0$ to $\mathcal{B}_m$).
We define the relative size of $B_i^{(0)}$ (resp.\ $B_i^{(m)}$) as 
$S_i^{(0)} = \frac{|B_i^{(0)}|}{2^n}$ (resp. $S_i^{(m)} = \frac{|B_i^{(m)}|}{2^n}$).
From this, we derive $\vec{S}_0$ and $\vec{S}_m$, the vectors of dimension $k$ of 
relative sizes of the related attraction basins of $M_0$ and $M_m$, and we define the 
following index of relative size similarity by:
\begin{equation*}
    \Delta_{\vec{S}} = \Delta(\vec{S}_0, \vec{S}_m) =
        d_{L1}(\vec{S}_0, \vec{S}_m) = 
        \sum_{i = 1}^k \left| S_i^{(0)} - S_i^{(m)} \right|\text{.}
\end{equation*}

\paragraph{Step 3.2: Trajectorial structure similarity up to isomorphism.}

Let $\mathscr{G}_0$ and $\mathscr{G}_m$ be the transition graphs resulting from the 
parallel evolution of $M_0$ and $M_m$. 
Considering $G = (V, E)$ be a digraph, the \emph{reversed digraph} of $G$, 
denoted by $G^{\text{rev}} = (V, E^{\text{rev}})$, is the digraph such that 
$(i, j) \in E^{\text{rev}}$ if and only if $(j, i) \in E$.
Let $\mathscr{G}_0^{\text{rev}}$ and $\mathscr{G}_m^{\text{rev}}$ be the reversed  
digraphs of $\mathscr{G}_0$ and $\mathscr{G}_m$, respectively. 
Given $i \in \{1, \ldots, k\}$, we denote by $T_i^{\max}$ (resp. $T_i^{\min}$) 
the length of the longest (resp.\ shortest) path from $\mathcal{A}_i$ in 
$\mathcal{G}_0^{\text{rev}}$.
We derive $\vec{T}_0^{\max} = (T_1^{\max}, \ldots, T_k^{\max})$ and 
$\vec{T}_0^{\min} = (T_1^{\min}, \ldots, T_k^{\min})$ the vectors of maximal and 
minimal convergence times (respectively) of $M_0$ towards each of its relevant 
attractors. 
Similar definitions are given on $M_m$ for $\vec{T}_{m}^{\max}$ and 
$\vec{T}_{\,m}^{\min}$ by basing on $\mathscr{G}_m$.
From this, and $\vec{T}_0 = (\vec{T}_0^{\max},\vec{T}_0^{\min})$ and 
$\vec{T}_m = (\vec{T}_m^{\max},\vec{T}_m^{\min})$, we define the following 
trajectorial similarity index:
\begin{equation*}
    \Delta_{\vec{T}} = \Delta(\vec{T}_0,\vec{T}_m) = 
        d_{L1}(\vec{T}_0^{\max}, \vec{T}_m^{\max}) +
        d_{L1}(\vec{T}_0^{\min}, \vec{T}_m^{\min})\text{.}
\end{equation*}

\paragraph{Step 3.3: Trajectorial structure similarity.}

This step consists in measuring the level of non-conservation of original 
configurations in the associated attraction basins of $M^m$, thanks to the the 
following quantity: $\traj_m = 
\sum_{i = 1}^k \left| B_i^{(0)} \setminus B_i^{(m)} \right|$.

\paragraph{Step 3.4: Minimizing instabilities and frustrations}

Here, considering that every attractor is likely, as its attraction basin, we 
want to ensure the conservation of models having the most stable dynamics.
To do so, given a model $M_m$, we compute two sensitivity measures based on instabilities 
and frustrations as follows, for all $i \in \{1, \ldots, k\}$:
\begin{equation*}
    \Inst_{B_i^{(m)}} = 
        \frac{\sum_{x \in B_i^{(m)}} \Inst^{(m)}(x)}{\left| B_i^{(m)} \right|}
    \quad\text{and}\quad
    \Frus_{B_i^{(m)}} = 
        \frac{\sum_{x \in B_i^{(m)}} \Frus^{(m)}(x)}{\left| B_i^{(m)} \right|}\text{.}
\end{equation*}
From this, we define the global sensitivity index for $M_m$ such that:
\begin{equation*}
    \sens_m = 
    \sum_{i=1}^k \left( \Inst_{B_i^{(m)}} + \Frus_{B_i^{(m)}} \right) \text{.}
\end{equation*}

\section{Application to osteogenesis}
\label{sec:appli}

\subsection{Definition of $M_0$}

This work uses a biological model of the osteogenesis regulation network as a case 
study. 
Specifically, we focus on the \tgfb{} subnetwork, which represents the core of the 
genetic regulation involved in osteogenesis imperfecta~\cite{demongeot2025boolean}. 
Osteogenesis is a highly regulated process involving multiple signaling pathways and 
transcription factors.

The reference model, $M_0$, is derived from a systemic approximation where the 
interaction digraph encodes experimentally observed activations and inhibitions. 
Previous analyses~\cite{demongeot2025boolean} indicate that, under the parallel 
update mode, this subnetwork exhibits a multi-stable landscape. 
While a small fraction of configurations converge toward a fixed point representing 
physiological osteogenesis, the majority evolve toward pathological attractors 
(characterized by inactive \tgfb{}) or spurious limit cycles without known 
phenotypic correlates.

\begin{table}[t!]
  \centerline{
    \begin{tabular}{lccccc}
      \hline
      Node & Attractor 1 & Attractor 2 & Attractor 3 & Attractor 4 & Attractor 5 \\
      \hline\hline
      \miR{}   & \zero & \one & \one & \one\one\one & \one\one\one \\
      \zeb{}   & \one & \zero & \zero & \zero\zero\zero & \zero\zero\zero \\
      \spone{} & \zero & \one & \one & \one\one\one & \one\one\one \\
      \smad{}  & \one & \zero & \zero & \zero\zero\zero & \zero\zero\zero \\
      \tgfb{}  & \one & \zero & \zero & \zero\zero\zero & \zero\zero\zero \\
      \notch{} & \zero & \zero & \one & \one\zero\zero & \one\one\zero \\
      \col{}   & \one & \zero & \zero & \zero\zero\zero & \zero\zero\zero \\
      \ccnd{}  & \zero & \zero & \one & \zero\zero\one & \one\zero\one \\
      \pax{}   & \zero & \zero & \one & \zero\one\zero & \zero\one\one \\
      \hline\hline
      Basin Size & 32 & 182 & 8 & 218 & 72 \\
      \hline
    \end{tabular}
  }\bigskip
  \caption{Attractors of the osteogenesis regulation network.}
  \label{tbl2}
\end{table}

\begin{figure}[t!]
	\centerline{
	   \includegraphics[width=.8\textwidth]{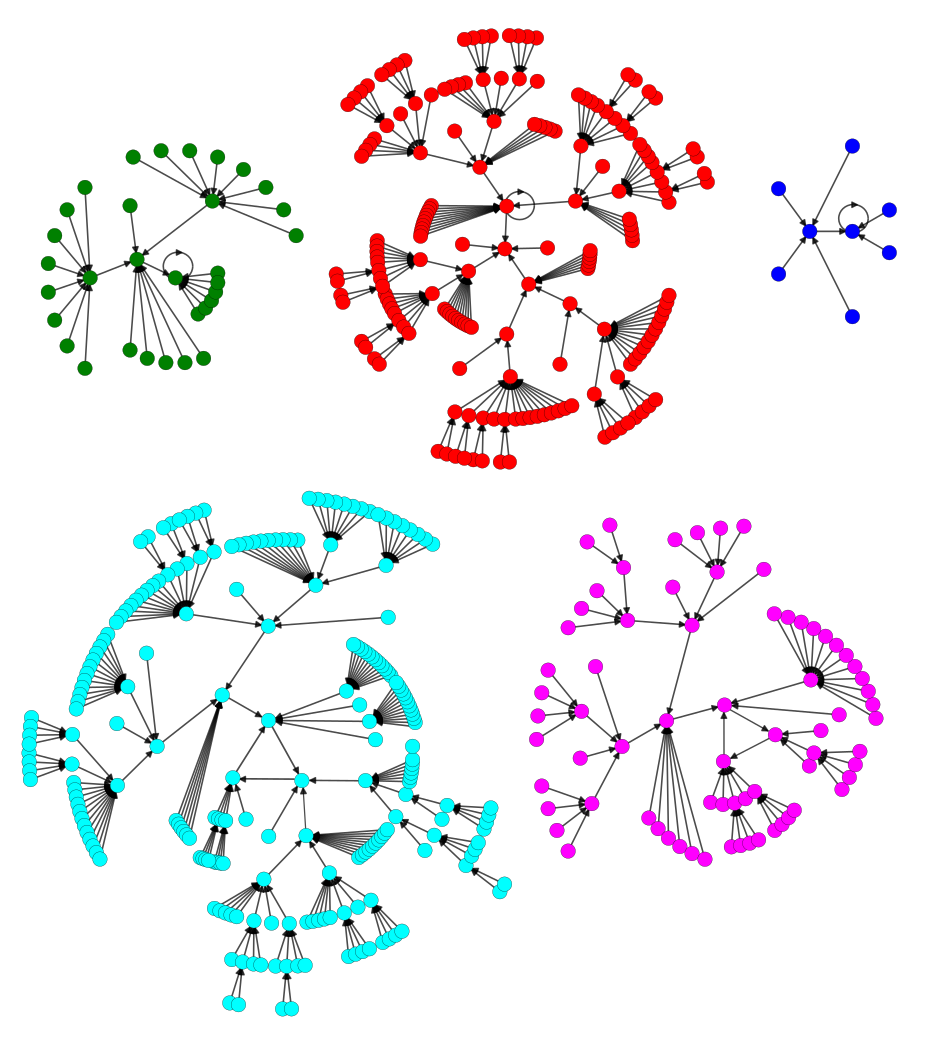}
    }
	\caption{Attraction basins of the osteogenesis network under the parallel update 
    mode. 
    For the 9-node network, attractors 1, 2, and 3 are fixed points, with basin sizes 
    of 32, 182 and 8 configurations, respectively. 
    Attractors 4 and 5 are limit cycles of length 3, with basin sizes of 218 and 72 
    configurations, respectively. 
    Colors indicate the corresponding basins: green (attractor 1),  (attractor 2), 
    blue (attractor 3), cyan (attractor 4), and pink (attractor 5).}
	\label{FIG:AG}
\end{figure}

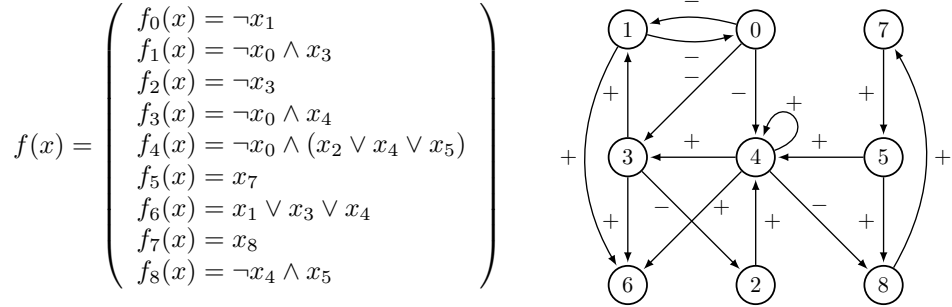
\begin{figure}[t!]
	\centerline{
		\begin{minipage}{.45\textwidth}
			\centerline{$f(x) = \left(\begin{array}{l}
				f_0(x) = \neg x_1\\
				f_1(x) = \neg x_0 \land x_3\\
				f_2(x) = \neg x_3\\
				f_3(x) = \neg x_0 \land x_4\\
				f_4(x) = \neg x_0 \land (x_2 \lor x_4 \lor x_5)\\
				f_5(x) = x_7\\
				f_6(x) = x_1 \lor x_3 \lor x_4\\
				f_7(x) = x_8\\
				f_8(x) = \neg x_4 \land x_5
			\end{array}\right)$}
		\end{minipage}
		\hfill
		\begin{minipage}{.42\textwidth}
			\centerline{\scalebox{0.85}{\begin{tikzpicture}[>=latex,auto]
				\tikzstyle{node} = [circle, thick, draw, minimum size=6mm, inner 
                sep=1pt]
				
				% Disposición en cuadrícula estructurada (3 columnas, 3 filas)
				% Fila superior
				\node[node](n2) at (0, 0) {$1$};
				\node[node](n1) at (2, 0) {$0$};
				\node[node](n8) at (4, 0) {$7$};
				
				% Fila central
				\node[node](n4) at (0, -2) {$3$};
				\node[node](n5) at (2, -2) {$4$};
				\node[node](n6) at (4, -2) {$5$};
				
				% Fila inferior
				\node[node](n7) at (0, -4) {$6$};
				\node[node](n3) at (2, -4) {$2$};
				\node[node](n9) at (4, -4) {$8$};

				% Aristas y regulaciones
				% Interacciones entre 1 y 2
				\draw[thick, ->] (n2) edge [bend right=15] node [swap] {$-$} (n1);
				\draw[thick, ->] (n1) edge [bend right=15] node [swap] {$-$} (n2);
				
				% Regulaciones de SMAD3 (4)
				\draw[thick, ->] (n4) edge node {$+$} (n2);
				\draw[thick, ->] (n4) edge node [pos=0.2, below] {$-$} (n3);
				\draw[thick, ->] (n4) edge node [swap] {$+$} (n7);
				
				% Regulaciones de miR200 (1) a otros
				\draw[thick, ->] (n1) edge node [above] {$-$} (n4);
				\draw[thick, ->] (n1) edge node [left] {$-$} (n5);
				
				% Regulaciones hacia TGFB (5) y desde TGFB
				\draw[thick, ->] (n5) edge node [swap] {$+$} (n4);
				\draw[thick, ->] (n3) edge node [right] {$+$} (n5);
				\draw[thick, ->] (n6) edge node [swap] {$+$} (n5);
				%\draw[thick, ->] (n5) edge [loop right] node {$+$} (n5);
                \draw[thick, ->] (n5) edge [loop, out=20, in=70, looseness=8] node [above] {$+$} (n5);
				\draw[thick, ->] (n5) edge node [above] {$-$} (n9);
				
				% Regulaciones del sumidero Col1A1 (7)
				\draw[thick, ->] (n2) edge [bend right=30] node [swap] {$+$} (n7);
				\draw[thick, ->] (n5) edge node [pos=0.2, below] {$+$} (n7);
				
				% Eje de Notch1 (6), CCND1 (8) y PAX8 (9)
				\draw[thick, ->] (n8) edge node [swap] {$+$} (n6);
				\draw[thick, ->] (n6) edge node [swap] {$+$} (n9);
				\draw[thick, ->] (n9) edge [bend right=30] node [swap] {$+$} (n8);
			\end{tikzpicture}}}
		\end{minipage}
    }
    \caption{Formal representation of the $M_0$ BN for the \tgfb{} osteogenesis 
        subnetwork. 
        (Left panel) BN $f$ associated with $M_0$. 
        (Right panel) Interaction digraph $G_f$, where nodes represent genetic 
        components and arcs denote regulatory interactions; positive and negative 
        signs indicate activation and inhibition, respectively.
        The list of components corresponds to the following order-preserving sequence 
        of biological genes: 
        \miR{}, \zeb{}, \spone{}, \smad{}, \tgfb{}, \notch{}, \col{}, \ccnd{}, \pax{}.}
    \label{fig:Osteo9_grid}
\end{figure}

Structurally, $M_0$ is defined as a 9-node BN, as illustrated in 
Figure~\ref{fig:Osteo9_grid}, which is the osteogenesis regulation subnetwork. 
A structural assessment of the associated interaction digraph reveals a complex 
topology comprising $5$ positive cycles and $2$ negative cycles. 
The attractors under the parallel update mode, along with their respective basin sizes 
and relative frequencies, are summarized in Table~\ref{tbl2} and a representation of 
its underlying transition graph is depicted in Figure~\ref{FIG:AG}. 
With a baseline sensitivity index of $51.89$, these properties establish the starting 
point for the application of our parsimonious protocol.

\subsection{Model space exploration and incremental pruning}

The inference process is constrained to eliminate spurious limit cycles, requiring 
candidate models to converge exclusively toward the biologically relevant fixed 
points. 
Exploring the syntactic search space for the $9$ nodes of $M_0$ (considering up to 
three literals per function) yields a total of 51,138 potential models. 

An initial filtering consisting in conserving only the models admitting the expected 
fixed points leads to 480 candidate models. 
These candidate models have been shown to be highly heterogeneous in their settings, 
and differ on six components out of nine: \pax, \ccnd, \notch, \tgfb{}, \miR, and 
\zeb{}.
%Some components emerged as rigid dynamical constraints: no valid replacement functions 
%were identified for \smad{}, \spone{}, and Col1A1. 
%Local logical alterations to these components consistently destabilized the desired 
%fixed points or generated spurious attractors. 
%Consequently, the 480 valid candidates were concentrated within the remaining nodes 
%(PAX8, CCND1, \notch, \tgfb{}, \miR, and \zeb{}), suggesting that while the global 
%dynamics are sensitive, specific topological loci are more permissive to logical 
%interventions.

To identify the most consistent models with the reference system $M_0$, an incremental 
pruning protocol is applied. 
At each stage, a $P = 20\%$ retention target is enforced based on structural and 
dynamical similarity indices. 
To ensure scientific rigor and avoid arbitrary exclusions, a tie-handling logic is 
implemented: 
the pruning threshold is defined by the value at the 20th percentile, but all 
candidate models sharing that value are preserved. 
This approach accounts for the discrete nature of the logical search space, where 
multiple models often exhibit identical dynamical properties. 
The refinement process is summarized below:
\begin{itemize}
\item Structural similarity ($\Delta_G$): 
    Because of ties, we keep $21.67\%$ of the models, which corresponds to a value for 
    $\Delta_G$ less than or equal to $2.0$.
    This first criterion leads to the conservation of $104$ models. 
\item Basin size organization ($\Delta_{\vec{S}}$): 
    Due to a high density of models with identical basin distributions, the effective 
    retention rate is $69.23\%$, which corresponds to a value for $\Delta_{\vec{S}}$ 
    less than or equal to $0.566$.
    This criterion leads to the conservation of $72$ models. 
\item Convergence time ($\Delta_{\vec{T}}$): 
    Because of ties, we keep $37.50\%$ of the models, which corresponds to a value for 
    $\Delta_{\vec{T}}$ less than or equal to $2.0$.
    This criterion leads to the conservation of $27$ models.
\item Trajectorial isomorphism ($\traj$): 
    Since the distance between the majority of surviving models and $M_0$ equals 
    $0.0$, we keep $81.48\%$ of the models. 
    This criterion leads to the conservation of $22$ models. 
\item Sensitivity optimization ($\sens$): 
    This final step leads to keep $27.27\%$ of the models, which corresponds to a 
    value for $\sens$ less than or equal to $27.75$.
    This criterion leads to the conservation of $6$ models at the end. 
\end{itemize}

These six models represent the most robust logical alternatives that preserve the 
essential osteogenesis attractors while maintaining a high degree of dynamical 
isomorphism with the reference model $M_0$. 
These alternative models are presented in Table~\ref{tbl3}.

\begin{table}[t!]
    \centerline{\rotatebox{90}{\resizebox{1.36\linewidth}{!}{
        \begin{tabular}{l||l|l||c|c|c|c|c||c|c|c|c||c||c||c}
            \hline
            Node & Original function & Resulting function & $\pcycle$ & $\ncycle$ & 
                $\pcomp$ & $\ncomp$ & $\Delta G$ & Att. 1 & Att. 2 & Att. 3 & 
                $\Delta\vec{S}$ & $\Delta\vec{T}$ & $\traj$ & $\sens$ \\
            \hline\hline
            \ccnd{} & \pax{} & $\ccnd{} \land \pax{}$ & 6 & 2 & 7 & 6 & 1 & 32 & 472 & 
                8 & 0.566 & 1 & 0 & 26.368 \\
            \ccnd{} & \pax{} & $\notch{} \land \ccnd{} \land \pax{}$ & 7 & 2 & 7 & 6 & 
                2 & 32 & 472 & 8 & 0.566 & 1 & 0 & 27.629 \\
            \ccnd{} & \pax{}  & $\lnot \tgfb{} \land \ccnd{} \land \pax{}$ & 6 & 3 & 
                7 & 6 & 2 & 32 & 472 & 8 & 0.566 & 1 & 0 & 27.745 \\     
            \pax{} & $\lnot \tgfb{} \land \notch{}$ & 
                $\lnot \tgfb{} \land \notch{} \land \pax{}$ & 6 & 2 & 7 & 6 & 1 & 32 & 
                472 & 8 & 0.566 & 0 & 0 & 26.737 \\
            \notch{} & \ccnd{} & $\notch{} \land \ccnd{}$ & 6 & 2 & 7 & 6 & 1 & 32 & 
                472 & 8 & 0.566 & 2 & 0 & 26.368 \\
            \notch{} & \ccnd{} & $\lnot \tgfb{} \land \notch{} \land \ccnd{}$ & 6 & 
                3 & 7 & 6 & 2 & 32 & 472 & 8 & 0.566 & 1 & 0 & 27.745 \\
            
            \hline
        \end{tabular}
    }}}\medskip
    
    \caption{Models obtained as results of the pruning protocol.}
    \label{tbl3}
\end{table}

\subsection{Analysis of candidate models}

The function $f_{\notch{}}(x) = x_{\notch{}} \land x_{\ccnd{}}$ can be regarded as a 
plausible refinement of the original function, since the original network already 
relates \notch{} activation to a \ccnd-associated proliferative state, and the 
available literature further supports a functional \notch--\ccnd{} connection in 
skeletal cells~\cite{Tian2017NotchMSC,Zanotti2016NotchSkeleton}. 
In this context, the added self-loop may be interpreted as a persistence term, whereby 
\ccnd{} contributes to \notch{} activation, while pre-existing \notch{} helps sustain 
its expression.

Along the same lines, the function $f_{\notch{}}(x) = \lnot x_{\tgfb{}} \land 
x_{\notch{}} \land x_{\ccnd{}}$ is also consistent with the evidence discussed above. 
In the original model, reduced \tgfb{} activity indirectly relieves repression of 
\pax{}, thereby favoring \ccnd{} maintenance and, consequently, conditions compatible 
with sustained \notch{} activity~\cite{demongeot2025boolean}. 
However, there is no explicit evidence supporting direct transcriptional 
self-regulation of \notch{}.

For $f_{\pax{}}(x) = \lnot x_{\tgfb{}} \land x_{\notch{}} \land x_{\pax{}}$, the 
strongest support comes from the inhibitory effect of \tgfb{} on \pax{} activity. 
Experimental studies have shown that \tgfb{} reduces \pax{} expression, protein 
abundance, and DNA-binding activity through \smad{}-dependent mechanisms, supporting 
$\lnot x_{\tgfb{}}$ as a plausible permissive condition for \pax{} 
maintenance~\cite{Kang2001Pax8TGFB1,Costamagna2004Pax8Smad3}. 
In parallel, developmental evidence suggests that \notch{} signaling may positively 
influence \pax{} expression domains.
However, this observation is context-dependent and does not establish a general direct 
regulatory link with \pax{}~\cite{Jayasena2008NotchPax8}. 
Thus, this function should be regarded as a tentative or alternative formulation 
rather than a strongly supported regulatory relationship. 
The self-dependence term likely reflects maintenance of an already active 
transcriptional state rather than a strict self-regulation.

Finally, the functions $f_{\ccnd{}}(x) = x_{\ccnd{}} \land x_{\pax{}}$ and 
$f_{\ccnd{}}(x) = x_{\notch{}} \land x_{\ccnd{}} \land x_{\pax{}}$ can be interpreted 
within a common biological framework in which sustained \ccnd{} expression depends on 
a permissive \pax{}-positive background, with \notch{} providing an additional 
regulatory layer in the more restrictive formulation. 
This interpretation is biologically plausible because \pax{} has been implicated in 
osteogenic differentiation of human bone marrow mesenchymal stem cells, while studies 
in human cellular systems have shown that \pax{} can regulate an enhancer associated 
with \ccnd{} expression~\cite{Ni2025LncMSTRG25PAX8,Patel2022PAX8CCND1}. 
In parallel, \notch{} signaling has been related to \ccnd{} induction, maintenance of 
proliferative capacity, and regulation of early osteogenic progression in mesenchymal 
and skeletal cell systems~\cite{Tian2017NotchMSC,Zanotti2016NotchSkeleton}. 
These observations support a model in which \ccnd{} maintenance is compatible 
with a \pax{}-positive osteogenic state and may be further reinforced by active 
\notch{} signaling, even though the complete combined interaction has not yet been 
directly demonstrated in osteogenic cells. 
The function $f_{\ccnd{}}(x) = \lnot x_{\tgfb{}} \land x_{\ccnd{}} \land x_{\pax{}}$ 
is also consistent with the evidence discussed above, along the same lines. In the 
original model, reduced \tgfb{} activity indirectly relieves repression of \pax{}, 
thereby favoring \ccnd{} maintenance~\cite{demongeot2025boolean}.

\section{Conclusion}
\label{sec:conclu}

We proposed a parsimonious computational framework to explore alternative local 
functions in Boolean networks while preserving a prescribed set of biologically 
relevant attractors. 
In the osteogenesis case study, the method reduced a search space of 51,138 candidate 
models to only 6 robust alternatives that preserve the prescribed attractors while 
eliminating spurious asymptotic behaviors. 
Importantly, these 6 solutions involve rewritings of only three local functions, namely 
\ccnd{}, \notch{} and \pax{}. 
Beyond model selection, this shows that the approach can help pinpoint specific regions 
of the reference network that are likely to be suboptimal from a dynamical viewpoint. 
In that sense, the framework can also be used as a practical aid during the modelling 
process, by guiding the modeler toward the components whose logical rules deserve 
revision.

The ordering of the pruning steps was chosen primarily for biological reasons, giving 
priority to the preservation of $\mathcal{A}^*$ before applying additional structural 
and dynamical similarity criteria. 
For larger Boolean networks, however, computational scalability may require reordering 
some stages of the protocol, in particular by using $\Delta_G$ as a faster preliminary 
filter before full attractor verification, notably for BNs of size greater than $15$ 
nodes.

Overall, the strong reduction achieved by the incremental pruning procedure shows that 
this approach can identify a small family of biologically plausible candidate models 
from a very large logical search space. 
These final networks should nevertheless be interpreted as testable hypotheses, 
especially regarding the inferred self-regulatory terms, which still require 
biological validation. 
%This validation implies the consideration of all possible interactions between genes 
%involved in the studied regulatory network, as given with a weighting of their 
%biological relevance in the \textsc{signor} database~\cite{losurdo_signor}.
As a natural direction for future work, it would be valuable to extend the analysis to 
alternative update schemes, particularly deterministic asynchronous block-sequential 
dynamics, in order to evaluate the stability of the selected networks under 
perturbations and to assess the robustness of their asymptotic behavior beyond the 
synchronous setting.

Eventually, we notice that in the reference model $M_0$, the attraction basin  of the 
non-pathological state remains comparatively small. 
Under a uniform sampling of initial configurations, this would make pathological 
outcomes more likely, which points to a limitation of the original dynamical landscape. 
This suggests that preserving the target attractors is not the only relevant 
requirement, and that the relative basin sizes should also be taken into account when 
assessing biological plausibility. 
In this perspective, it would be interesting to adapt the criterion $\Delta_S$ so as to 
give more weight to models whose dynamical landscape is more favorable to the absence 
of pathology.

%Discussion on the order of the steps, saying the order has been chosen according to 
%biological reasons. For instance, for BNs of sizes greater than $15$, 
%it would be necessary, from an algorithmic and complexity perspective, to permute the 
%step on $\mathcal{A}^*$ and $\Delta(G_0, G_m)$. 

\paragraph{Acknowledgements}
The authors received support from the following projects: 
HORIZON-MSCA-2022-SE-01101131549 ACANCOS (AER, EG, SS, LT), 
ANID-MILENIO-NCN2024\_047 (EG), ANID-MILENIO-NCN2024\_103 (EG), 
ANID FONDECYT 1250984 regular (EG, MMM), 
ANR-24-CE48-7504 ALARICE (SS), and
ANR-24-CE45-1164 MITOMETATIS (LT).

%\bibliographystyle{plain}
%\bibliography{biblio.bib}

\end{document}